\documentclass[aip,reprint,groupedaddress,floatfix]{revtex4-1}
\usepackage{amssymb,amsmath}
\usepackage{graphicx}
\graphicspath{{Figures/}}
\DeclareMathOperator{\sech}{sech}
\newcommand{\abs}[1]{\left\lvert#1\right\rvert}
\newcommand{\phiK}{\phi_{\rm{K}}}
\newcommand{\phiKbar}{\phi_{\bar{\rm{K}}}}
\newcommand{\KK}{K\={K}}
\renewcommand{\phi}{\varphi}
\newcommand{\vc}{v_{\rm c}}
\newcommand{\vin}{v_{\rm in}}
\newcommand{\vout}{v_{\rm out}}
\newcommand{\cm}{{\rm{cm}}}
\newcommand{\CC}{\mathcal{C}}
\newcommand{\XS}{X_{\rm S}}
\newcommand{\cH}{\mathcal{H}}

\begin{document}

\title{A Mechanical Analog of the Two-bounce Resonance of Solitary Waves: Modeling and Experiment}
\author{Roy H. Goodman}
\email{goodman@njit.edu}
\author{Aminur Rahman}
\email{ar276@njit.edu}
\author{Michael J. Bellanich}
\email{mbellanich@gmail.com}
\affiliation{Department of Mathematical Sciences, New Jersey Institute of Technology, Newark, NJ 07102}
\author{Catherine N. Morrison}
\email{cmorrison@veritasnj.org}
\affiliation{Veritas Christian Academy, Sparta Township, NJ 07871}

\date{\today}                                       

\begin{abstract}
We describe a simple mechanical system, a ball rolling along a specially-designed landscape, that mimics the dynamics of a well known phenomenon, the two-bounce resonance of solitary wave collisions, that has been seen in countless numerical simulations but never in the laboratory. We provide a brief history of the solitary wave problem, stressing the fundamental role collective-coordinate models played in understanding this phenomenon. We derive the  equations governing the motion of a point particle confined to such a surface and then design a surface on which to roll the ball, such that its motion will evolve under the same equations that approximately govern solitary wave collisions. We report on physical experiments, carried out in an undergraduate applied mathematics course, that seem to verify one aspect of chaotic scattering, the so-called two-bounce resonance.
\end{abstract}
%applications of chaos
\maketitle

\textbf{%
Solitary waves are solutions to partial differential equations that maintain their spatial profile while moving at constant speed. A wide variety of systems display a behavior called chaotic scattering when two such waves collide. The two waves may bounce off each other one or more times before escaping to infinity, or they may capture each other and never escape. The number of collisions and the final speed of those that separate depends in a complex way on their initial speeds. To our knowledge, this process has never been observed in a laboratory experiment involving solitary waves. The problem is described well by a small finite-dimensional system of ordinary differential equations. We describe an experiment, a ball rolling on a specially designed surface, that obeys the same system of ODE.  We report on laboratory experiments demonstrating that the experimental system has similar dynamics to the solitary wave collisions.
}
%\listoftodos

\section{Introduction}

\emph{Solitary waves}, localized structures that translate at constant velocity while maintaining their spatial profile, are among the most important concept in nonlinear physics. \emph{Solitons} are a class of solitary waves which possess an additional underlying mathematical structure, one which renders the equations solvable, and which, in particular, causes the behavior of colliding solitons to be strikingly simple: solitons survive a collision with their shape and velocity intact, but with their positions shifted by a finite, computable, amount.
 
\begin{figure}
   \centering
   \includegraphics[width=.9\columnwidth]{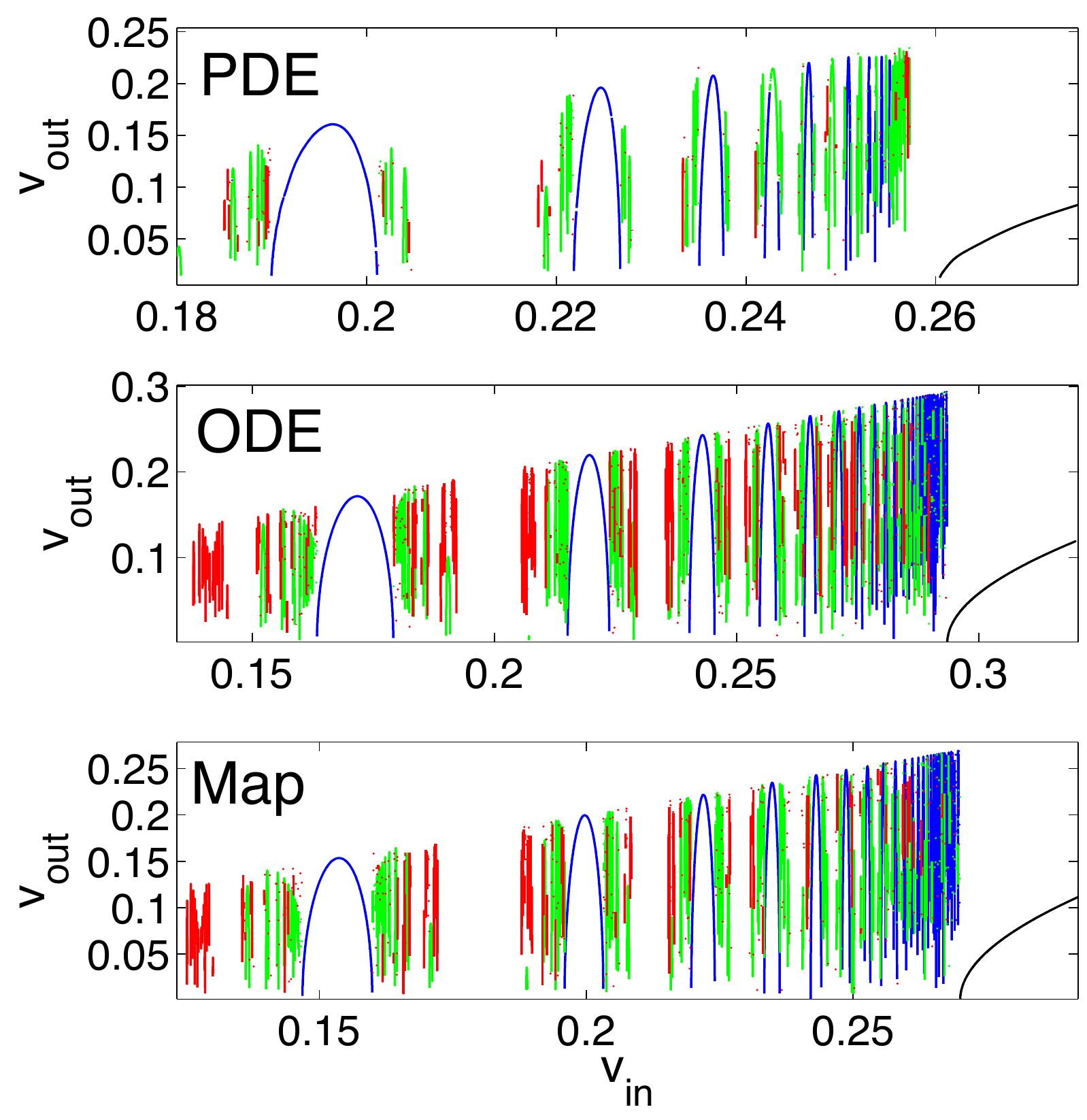} 
   \caption{The escape speed as a function of the input speed for a kink-antikink collision in the $\phi^4$ system for PDE~\eqref{phi4}, phenomenological ODE system~\eqref{CCODE}, and the discrete map~\eqref{themap}. Color indicates number of collisions before separation: one (black), two (blue), three (green), and four (red).}
\label{fig:inout}
\end{figure}

As long as scientists have known about solitary waves and had computers capable of simulating them, we have been colliding solitary waves together numerically and observing what comes out. An important example is the collision of so-called \emph{kink} and \emph{antikink} solutions to the $\phi^4$ equation,
\begin{equation}
\phi_{tt} - \phi_{xx} +\phi - \phi^3 = 0.
\label{phi4}
\end{equation}
The first studies in the 1970's gave a fleeting hint of rich structure and the first step toward understanding it. This  was explored more deeply and given a name, the \emph{two-bounce resonance}, in the 1980's. In the 1990's, more detailed numerical experiments revealed chaotic scattering. Finally, in the early 2000's, the mechanism behind this chaotic scattering was explained fully using techniques from dynamical systems. Figure~\ref{fig:inout} shows the speed at which a kink and antikink separate as a function their speed of approach, demonstrating a rich structure. Numerical simulations of partial differential equations (PDE) arising in diverse areas of physics have revealed the same phenomenon. In all that time, however, \emph{this behavior has never been reported in a physical experiment in a real nonlinear wave system}.

This paper describes a simple experiment---a ball set rolling on a manufactured landscape---which mimics the dynamics of solitary wave collisions and reproduces some features of Figure~\ref{fig:inout}.  The main tool for understanding solitary wave interaction has been the derivation and analysis of \emph{collective coordinate} (CC) models. The landscape is constructed such that the ball's equations of motion strongly resemble the ordinary differential equation system (ODE) governing the evolution of solution parameters described by a CC reduction describing solitary wave collisions in the $\phi^4$ equation.

Chaotic scattering is an appealing phenomenon to study because images such as Figure~\ref{fig:inout} cry out for explanation. The mathematical structure that has been developed is equally appealing, as is well-described by Ott.\cite{Ott:1993vs,Ott:2002}  A scattering process is a physical phenomenon in which an object's trajectory begins and ends in free motion (i.e.\ with zero acceleration) but spends a finite time in a region where it is subject to forces.  Such a process is called chaotic scattering if the final state depends in a complex, i.e.\ fractal or multi-fractal, manner on its initial state.

The paper is organized as follows.  
Section~\ref{sec:history} discusses where the problem comes from and what approaches have been most useful in understanding it. It begins with a brief recap of the history of solitons, followed by an explanation of the difference between solitons and garden-variety solitary waves. It then describes the long history of chaotic scattering of solitary waves, the progress that has been made, and that methods that have been used to explain the observations.
In Section~\ref{sec:model}, we describe the laboratory experiment and the mathematical model that describes it, while in Section~\ref{sec:experiment}, we present the experimental results, which we interpret as the two-bounce resonance, the aspect of chaotic scattering that is most robust in the presence of dissipation.
We end in Section~\ref{sec:plea} by reviewing the physical systems that have motivated many of earlier studies of the two-bounce/chaotic scattering phenomenon and expressing our hope that someone may perform this experiment to display this phenomenon in an extended physical system. 

\section{Historical motivation}
\label{sec:history}

\subsection*{Solitary waves and solitons}

\emph{Solitary waves} are solutions to a nonlinear evolutionary partial differential equation (PDE) that translate at a constant speed while maintaining its spatial profile.  \emph{Solitons} are solitary waves which exhibit unexpectedly simple dynamics upon collision. This is due to a hidden mathematical structure that was discovered beginning in the 1960's, first for the Korteweg-de Vries equation\cite{Zabusky:1965bb,Gardner:1967vy},
$$
u_t + 6 u u_x + u_{xxx} = 0,
$$
and later for many other equations, such as the sine-Gordon equation,
\begin{equation}
u_{tt}-u_{xx} + \sin{u} = 0.
\label{sG}
\end{equation}
Such soliton equations can be solved exactly using a method called the inverse scattering transform (IST).\cite{Ablowitz:1973ub} Good brief histories of solitary waves and solitons can be found elsewhere, e.g. in encyclopedia articles by Scott\cite{Scott:2004}, and by Zabusky and Porter.\cite{Zabusky:2010} 

The IST can be used to explicitly show that after colliding,  solitons continue propagating with the same shape and speed, but with a phase shift and a time delay.
Most nonlinear wave equations do not possess the IST, however, and solitary waves collisions in such systems display more complicated dynamicsn. Beginning in the 1970's, numerical simulations hinted this picture, but it took the development of more powerful computer hardware and numerical methods for the true complexity of this dynamics to fully emerge. The simultaneous development of \emph{collective coordinate methods} would prove useful for untangling this complexity.

\subsection*{Collective coordinates methods}

The term ``collective coordinates'' refers to a variety of  methods that are used to obtain simple ODE models that approximate the behavior of a PDE, at least as long as the solution stays in some small region of solution space.  A very useful approach is the ``variational method'', which applies to partial differential equations that arise as the Euler-Lagrange equations for a system with Lagrangian density
\begin{equation}
I= \iint {\mathcal L}(\phi,x,t) dx\, dt
\label{action}
\end{equation}
For example, solutions of the $\phi^4$ equation~\eqref{phi4} minimize the action due to
$$
{\mathcal L} = \frac{1}{2} \phi_t^2 - \frac{1}{2} \phi_x^2 + \frac{1}{2} \phi^2 -  \frac{1}{4} \phi^4
$$
over all $C^1$ function satisfying appropriate boundary conditions at infinity.

The variational method, due originally to Bondeson,\cite{Bondeson:1979bh}  works by constructing a solution ansatz whose spatial profile depends on a small number of time-dependent parameters and minimizing the action~\eqref{action} with respect to these parameters. The resulting equation is a finite-dimensional Lagrangian system of ODE governing the parameters' evolution. Many examples and some generalizations are given in the review paper of Malomed.\cite{Malomed:2002}

\subsection*{Numerical and analytical studies of solitary wave collisions}
Solitary waves exist in the $\phi^4$ equation~\eqref{phi4}, which is non-integrable, despite its clear similarity to the sine-Gordon equation~\eqref{sG}. These ``kink'' waves are given by
\begin{multline}
\label{phi4kink}
\phiK(x,t;v) =  \tanh{\frac{\xi}{\sqrt{2}}} \text{, where } \xi=\frac{x - v t}{\sqrt{1-v^2}} 
\\ \text{ for any } -1<v<1.
\end{multline}
The ``antikink'' solution is just $\phiKbar (x,t,v)=-\phiK(x,t;v)$.

 Many papers have been written about kink-antikink (\KK) collisions in this system. 
In 1975, Kudryavtsev\cite{Kudryavtsev:1975tc} found numerically that when $v=0.1$, the \KK\ pair coalesce into a single bound state at the origin which oscillates irregularly and is slowly damped to zero. He explained this with a formal argument based on potential energy which can be considered a first step toward an explanatory collective coordinates model.

In 1979, Sugiyama\cite{Sugiyama:1979gl} performed additional experiments for a few values of $v$ between 0.1 and 0.6. He found that those with speeds below a critical value $\vc\approx 0.25$ the \KK\ pair merge into a localized bound state, while for $v>\vc$, the pair collides inelastically. Careful  examination of his simulations showed that, upon collision, an oscillatory mode (sometimes called a ``shape mode'', since its effect is to alter the shape of the kink) is excited, which removes energy from the translation component. If the initial kinetic energy is less than the amount lost to the oscillatory mode, the \KK\ pair can not escape to infinity. If the $v>\vc$, they will escape to infinity but with reduced speed and with  oscillations superimposed. He also derived the first CC model for this phenomenon, and used it in a formal calculation to determine~$\vc$ in agreement with his numerical observations.

\begin{figure} 
   \centering
   \includegraphics[width=\columnwidth]{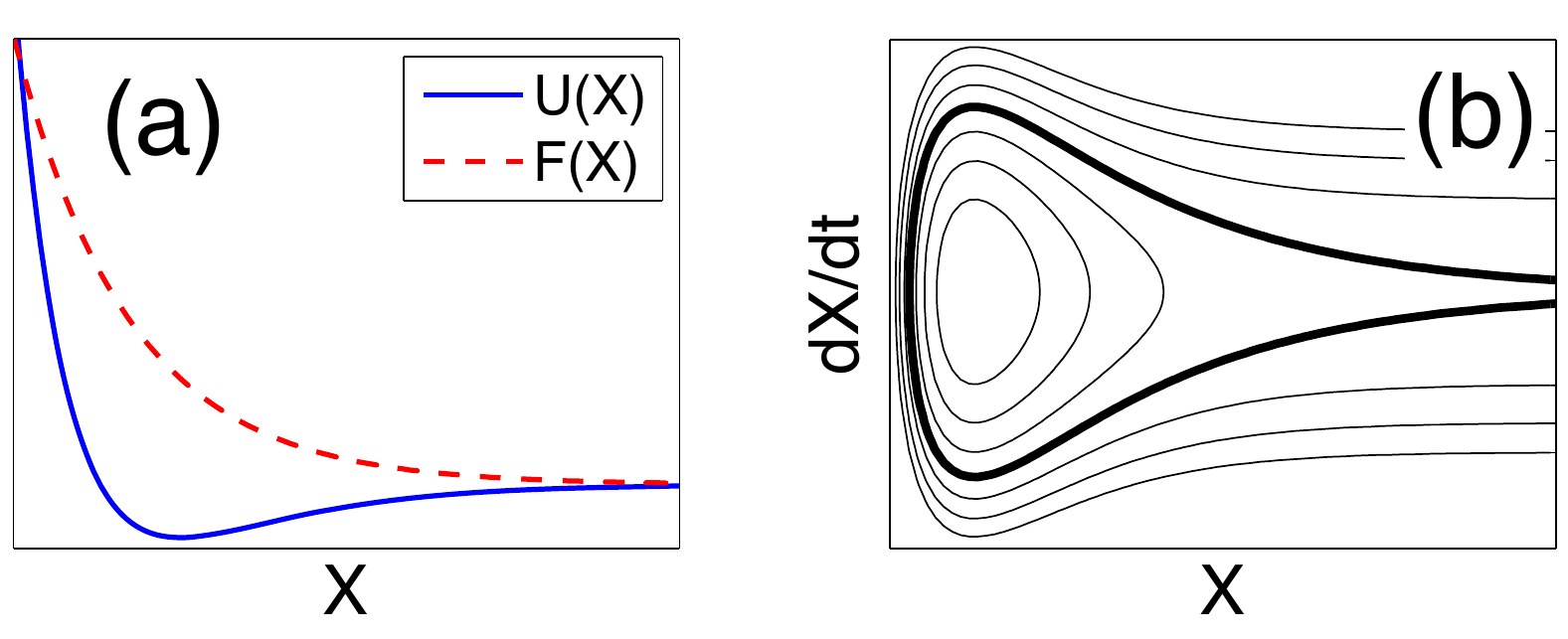} 
   \caption{(a) The potential and coupling function of equation~\eqref{CCODE}. (b)~The phase plane of the $X-\dot{X}$ subsystem, separatrix given by thick line.}
\label{fig:phaseplane}
\end{figure}

The shape mode is an eigenfunction for the linearization of equation~\eqref{phi4} about the kink solution~\eqref{phi4kink}, and is given by 
$$
\chi_1{(\xi)} = {\left(\frac{3}{\sqrt{2}}\right)}^{\tfrac{1}{2}} \sech{\frac{\xi}{\sqrt{2}}} \tanh{\frac{\xi}{\sqrt{2}}}
$$
and oscillates with frequency $\omega_1 = \sqrt{\tfrac{3}{2}}$. Sugiyama based his CC ODE for this system on the ansatz 
\begin{equation}
\label{ansatz}
\begin{split}
\phi(x,t) = & \phiK\left(x-X(t)\right) + \phiKbar\left(x+X(t)\right) - 1 \\
&+ A(t) \chi_1\left(x-X(t)\right) -A(t)\chi_1\left(x+X(t)\right).
\end{split}
\end{equation}
It leads to a two degree-of-freedom Hamiltonian system for the evolution of $X(t)$ and $A(t)$. The ODE system that arises is somewhat complicated,\cite{Sugiyama:1979gl} but its fundamental features are captured by the simpler model system
\begin{subequations}
\begin{align}
\ddot{X}(t) + U'(X) + c A F'(X) &=0\label{Xddot}; \\
\ddot{A}(t) + \omega^2 A + c F(X) &=  0\label{Addot},
\end{align}
\label{CCODE}
\end{subequations}
where $U(X) = e^{-2X}-e^{-X}$ and $F(X) = e^{-X}$, plotted in Figure~\ref{fig:phaseplane}(a). This conserves an energy 
$$
E = \frac{1}{2}\dot{X}^2 + U(X) + \frac{1}{2} \left( \dot{A}^2+ \omega^2 A^2 \right) + c A F(X).
$$
The term $U(X)$ is a potential energy describing the interaction of the two kinks, and was essentially derived by Kudryavtsev.\cite{Kudryavtsev:1975tc} For large $X$, both $F'(X)$ and $U'(X)$ vanish, so that the kink and antikink centers $\pm X(t)$ move at constant speed. The most important feature, for our purposes, is the phase diagram for $X$ when $c=0$, shown in Figure~\ref{fig:phaseplane}(b). The trajectories are level sets of the energy $E=\tfrac{1}{2}\dot{X}^2 + U(X)$, with $E<0$ on bounded orbits, a separatrix orbit with $E=0$ and unbounded orbits for $E>0$. We denote the separatrix orbit by $\XS(t)$, and note it is an even function of time. 

Figure~\ref{fig:inout} shows that this model qualitatively captures much of the dynamics of solitary wave collisions in~\eqref{phi4}. Observe, however, a few fundamental differences. While both systems conserve energy, in extended system~\eqref{phi4}, radiation (phonons) can carry energy away from the immediate vicinity of the kinks, effectively adding dissipation. This can be seen in the figure, where only solutions that escape after four or fewer collisions are plotted. In ODE~\eqref{CCODE}, all solutions escape except for a set with measure zero, while for the PDE, a significant fraction of solutions are trapped, due to energy loss to radiation. The radiation also steals some energy from solutions that do escape, so that $\vout<\vin$ even at the maxima of the resonance windows.

Also in 1979, Ablowitz, Kruskal, and Ladik reported on a series of numerical experiments for a few different nonlinear Klein-Gordon equations of the form $\phi_{tt} - \phi_{xx} + f(\phi) =0$, including the $\phi^4$ equation and the sine-Gordon equation.\cite{Ablowitz:1979uo}. Their numerical results were similar to Sugiyama's but the paper ends with an observation which was to prove important: On certain velocity intervals below $\vc$, the kink and antikink eventually separate, instead of forming a bound state as seen by  Kudryavtsev (from their References, it seems they had not seen Sugiyama's work at this point). When the initial velocity is 0.3---greater than $\vc$---the final velocity is 0.135. However, when the initial velocity is 0.2---less than $\vc$, the final velocity is $0.155$. They conclude ``The reason for the apparent `resonance' between these interacting aperiodic waves and the radiation is not yet fully understood.''  

In 1983, Campbell et al.\cite{Campbell:1983} performed a more systematic numerical sweep  of initial velocities and found significantly more detailed structure. Their calculation revealed the black curve and the leftmost nine blue curves in Figure~\ref{fig:inout}. The black curve shows the final velocities of all the solutions that escape after exactly one collision, i.e.\ those with initial speed above $\vc$, which they estimate numerically to be 0.26. The blue curve shows the final velocities of those with exactly two collisions. They called this phenomenon the \emph{two-bounce resonance}. 

\begin{figure}[tbp] 
   \includegraphics[width=.96\columnwidth]{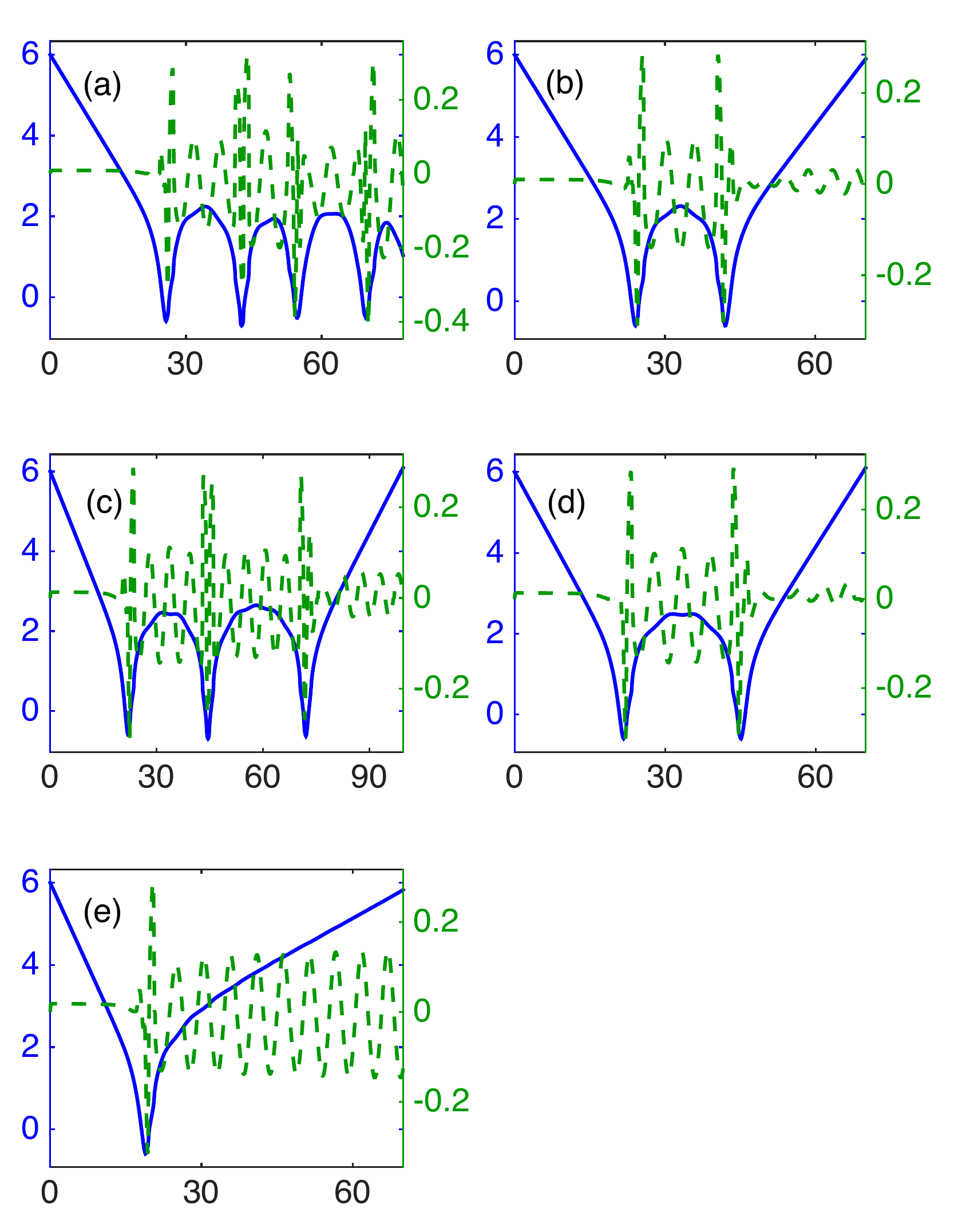} 
   \caption{Position $X(t)$  (solid, left $y$-axis) and shape-mode amplitude $A(t)$ (dashed, right $y$-axis) for selected initial velocities. (a) $v = 0.184$,  Capture. (b) $v =0.1986$, First two-bounce window. (c)~$v_0=0.2236$, a three-bounce window. (d)~$v=0.2268$, second two-bounce window. (e)~$v=0.27$ escape without capture. Vertical asymptote in (a) due to singularity of finite dimensional reduction, as per Caputo.\cite{Caputo:1991wt}}
\label{fig:PDE_X_a}
\end{figure}

The first question is what is the difference between solutions in one two-bounce window and those in the next. A good way to understand the solutions is to fit the numerical solution to the ansatz~\eqref{ansatz}, and plot the approximate values of $X(t)$ and $A(t)$. This is done in Figure~\ref{fig:PDE_X_a} for five increasing initial velocities. Subfigure~(a) shows a collision leading to capture; subfigures~(b) and~(d) show solutions from the first two two-bounce windows; (c)~shows a solution from a three-bounce window; (e)~shows a one bounce solution with $\vin>\vc$. The number of  oscillations of $A(t)$  is the same for all initial velocities in a given window and increases by one from one window to the next. Let $T_{\rm shape}$ be the period of the shape oscillation and $T_n$ be the time interval between the two collisions for the $\vin$ in the $n$th window at which $\vout$ is maximized, with $T_2$ and $T_3$ indicated in subfigures~(b) and~(d). We remark that the nonlinear projection~\eqref{ansatz} is singular when $X(t)=0$, because the coefficient of $A(t)$ then vanishes,\cite{Caputo:1991wt} in which case the fitting algorithm fails.

Treating the numerical results as laboratory data, Campbell et al.\ reasoned there should be a relation of the form 
\begin{equation}
T_n \approx  T_{\rm shape} \, n + \delta,
\label{eq:tn}
\end{equation} which they confirm by least-squares fitting the data to a line. They conjectured that at the first collision, the kink loses energy to the shape-mode oscillation, and that, if the timing is right on the second collision, the shape mode returns enough energy to the propagating mode to allow escape. What remained was to explain the mechanism.

A few years later, Anninos et al.\cite{Anninos:1991vo} showed, via more detailed experiments, the existence of higher bounce windows arranged in a fractal structure, i.e.\ of chaotic scattering, in the solitary wave collisions, including now, the three and four bounce windows of Figure~\ref{fig:inout}

Over the next two decades, these numerical experiments were reproduced for many other nonlinear wave systems, and Campbell et al.'s reasoning was applied to these systems as well. Campbell, as well as Peyrard, looked at a variety of $\phi^4$-like equations. They demonstrated that, at least for the systems they considered, the resonance phenomenon was only present when the linearization around the kink possesses an internal mode.\cite{Peyrard:1982,Peyrard:1983,Remoissenet:1984}
Of particular interest is a study by Fei et al.\cite{Fei:1992uu} which considers the sine-Gordon equation perturbed by a small localized defect,
\begin{equation}
\label{Fei}
u_{tt}-u_{xx}+\sin{u} = \epsilon \delta(x) \sin{u}.
\end{equation}
They find that kink solutions impinging on the defect show a behavior very similar to that seen in Figure~\ref{fig:inout}. They derive a CC system similar to system~\eqref{CCODE} where $A(t)$ now measures the amplitude of a mode localized in neighborhood of the defect. They use conservation of energy to derive an implicit formula for $\vc(\epsilon)$ and apply Campbell et al.'s reasoning to fit the time between collisions to the period of the secondary oscillator. 

\subsection*{Reduction of collective coordinates to an iterated map}

Beginning in 2004, Goodman and Haberman published a series of papers explaining the chaotic scattering in detail by reducing the CC ODE system to a discrete-time iterated map, called a separatrix map or scattering map.\cite{Zaslavsky:2007,Gidea:2014wx} This was first done\cite{Goodman:2004ca} for the system studied by Fei,\cite{Fei:1992uu} because the the CC ODE system for equation~\eqref{Fei} has a small parameter~$\epsilon$ that can be used in a perturbation analysis. Explicit formulas were found to approximate the critical velocities and the two- and three-bounce resonance, eliminating the need for data fitting as in equation~\eqref{eq:tn}. Over subsequent papers,\cite{Goodman:2005vp,Goodman:2005vv,Goodman:2007wj} the derivation of the map was streamlined and applied to other solitary waves systems and the model system~\eqref{CCODE}. It was subsequently put in a more explicit form.\cite{Goodman:2008gk} We outline this last approach here.

\begin{figure} 
   \centering
   \includegraphics[width=\columnwidth]{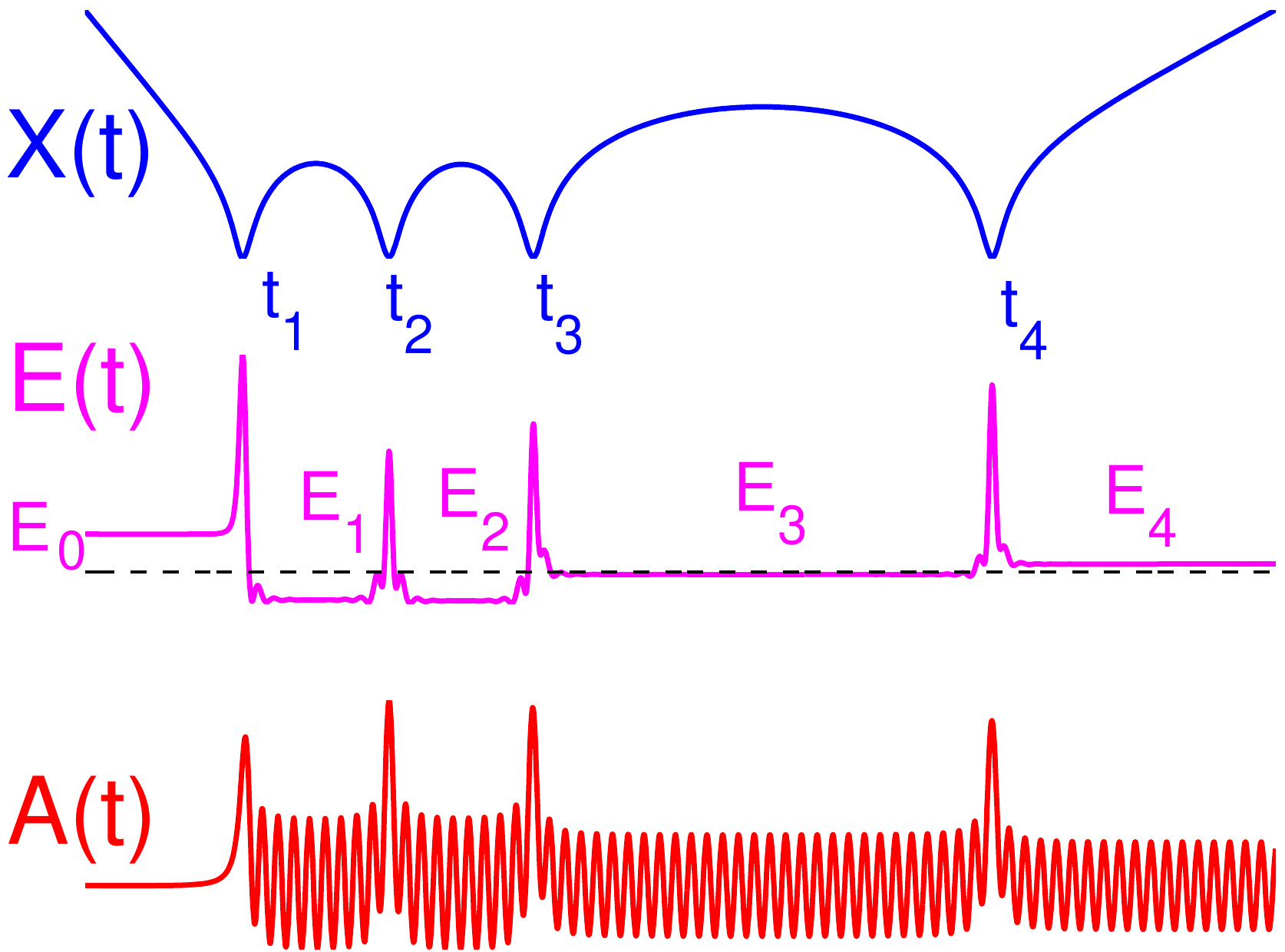} 
   \caption{Components used to construct the iterated map. Top $X(t)$ showing the collision times $t_j$. Middle: $E(X,\dot{X})$ the energy of the $X$-component. Bottom: $A(t)$.}
\label{fig:map_components}
\end{figure}

Figure~\ref{fig:map_components} depicts the results of one numerical simulation of the initial value problem~\eqref{CCODE} with the solitons initially far apart, propagating toward each other, and with the shape mode unexcited, i.e $X\gg1$, $\dot{X}<0$, and $A = \dot{A} = 0$. Additionally $\dot X$ is small enough for capture to occur. 
In this simulation, $E(t)>0$ before the collision, so that the solution is outside the separatrix in Figure~\ref{fig:phaseplane}(b), and crosses to the inside where $E<0$ at the first collision time $t_1$. At each subsequent collision $E(t)$ jumps, reaching a plateau $E_j$ between collisions at $t_{j-1}$ and $t_j$, and escaping to infinity when $E(t)>0$ once again. Upon each collision the amplitude and phase of $A(t)$ jump. We represent the solution before the collision at time $t_j$ by the energy level $E_j$ and by assuming  
\begin{equation}
A(t) \sim \CC \left( c_j \cos{\omega(t-t_j)} + s_j \sin{\omega(t-t_j)} \right)
\label{A_before}
\end{equation}
for $t$ between $t_{j-1}$ and $t_j$, where setting the constant 
$$
\CC = \frac{c}{\omega} \int_{-\infty}^{\infty} F(\XS(t)) \cos{\omega t} dt = 
\frac{c}{\omega} \int_{-\infty}^{\infty} F(\XS(t)) e^{i\omega t} dt
$$  
allows us to scale the variables to be $O(1)$ in the final form of the system.
We  seek a map $(E_{j+1},c_{j+1},s_{j+1}) = M(E_j,c_j,s_j)$. 

The map is constructed by building a matched asymptotic expansion which alternates between ``outer expansions'' where $X(t)$ is approximated by $\XS(t-t_j)$, Figure~\ref{fig:phaseplane}(b), and ``inner expansions''. On the outer solution, valid when $X(t)\gg1$, the modes exchange energy. We find, by variation of parameters, that
outer solutions with ``before-collision'' condition~\eqref{A_before} as $t-t_j \to -\infty$, satisfy the ``after-collision'' condition
\begin{multline}
A(t) \sim \CC \left( c_j \cos{\omega(t-t_j)} + (s_j-1) \sin{\omega(t-t_j)} \right) \\ \text{as } t-t_j \to + \infty,
\label{A_after}
\end{multline}
and that 
\begin{equation}
E_{j+1} = E_j + \CC^2\omega^2\left(-\frac{1}{2}+s_j\right)
\label{DeltaE}
\end{equation}

The change of energy is computed using a \emph{Melnikov integral}.\cite{Zaslavsky:2007}. Condition~\eqref{A_after} is written in terms of $(t-t_j)$, whereas the form of the map requires $(t-t_{j+1})$. We can compute the time between collisions by matching between two the outer expansions connected by an inner expansion, which gives $t_{j+1}-t_j = \alpha E_{j+1}^{-1/2} + o(1)$ for some $\alpha$ determined by matched asymptotics. This gives, approximately,
$$
\begin{bmatrix}
c_{j+1} \\ s_{j+1}
\end{bmatrix} 
=
\begin{bmatrix}
\cos{\theta_{j+1}} & \sin{\theta_{j+1}} \\
- \sin{\theta_{j+1}} & \cos{\theta_{j+1}}
\end{bmatrix}
\begin{bmatrix}
c_j \\ s_j-1
\end{bmatrix},
$$
where $\theta_{j+1} = \omega (t_{j+1}-t_j)$.

The discrete map possesses a conserved quantity $\cH = E_j + \tfrac{1}{2} \omega^2 \CC^2(c_j^2 + s_j^2)$ that allows us to eliminate $E_j$ from the system.  Defining a complex variable $z_j = c_j + i s_j$, the new map may be written
\begin{equation}
z_{j+1} = e^{-i \alpha{\left(\cH - \abs{z_j-1}^2\right)}^{-1/2}}(z_j -1).
\label{themap}
\end{equation}

 Figure~\ref{fig:inout} shows that the map reproduces the qualitative and many quantitative features of the ODE system and enables the determination of many of the features of the dynamics of system~\eqref{CCODE}. For example, in the experiments depicted in Figure~\ref{fig:inout}, the initial condition is $z_j=0$. Therefore, we may estimate the critical velocity: the solution is captured if $E_1<0$, i.e.\ if $v <\vc = \omega\CC$, making use of equation~\eqref{DeltaE}. The analysis also allows us to calculate approximate formulas for the centers of the two- and three bounce resonance windows, allowing the computation of the constants in equation~\eqref{eq:tn} without  data fitting.\cite{Goodman:2008gk}

\section{A Physical Model of the Collective Coordinate Equations}
\label{sec:model}

We wish to design a surface such that a ball confined to roll along that surface satisfies equations of motion similar to system~\eqref{CCODE}. We derive the evolution equations corresponding for motion along a general surface and then design a surface that gives the desired dynamics.

\subsection{The mathematical model}
\begin{figure}[ht] 
   \centering
   \includegraphics[width=.95\columnwidth]{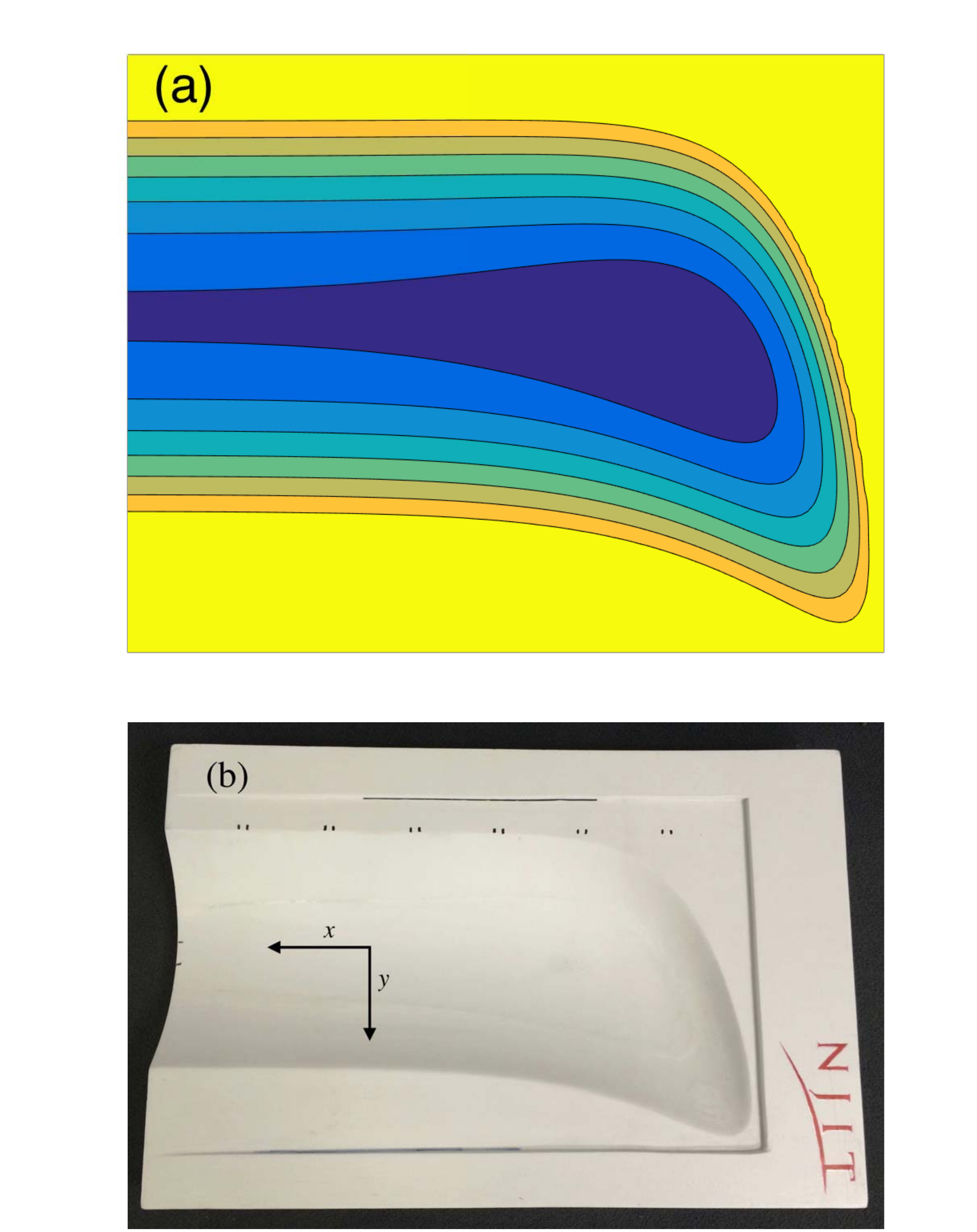}
      \caption{(a) Contour plot of the energy surface. (b) Photograph of the experimental surface.}
\label{fig:surface}
\end{figure}

Consider the behavior of a point  particle of mass $m$ confined to a surface $z=h(x,y)$ and moving under the influence of constant gravity in the $z$-direction, and ignoring any friction or other dissipative mechanisms.
We do not model in detail the rotation of the ball, instead noting that the rotational kinetic energy of a sphere rolling with speed $v$ is $\tfrac{2}{5}m v^2$. Adding this to the translational component gives a total kinetic energy
$$
T =  \frac{7}{5}\frac{m}{2} \left( \dot{x}^2 + \dot{y}^2 + \dot{z}^2 \right) =
 \frac{7}{5} \frac{m}{2} \left( 
\dot{x}^2 + \dot{y}^2 +{\left(h_x \dot{x} + h_y \dot{y}\right)}^2
 \right).
 $$
The gravitational potential energy is just
 $
 U = m g z = m g h(x,y).
 $
The evolution of a Lagrangian system with coordinates $q_j$ is governed by the Euler-Lagrange equations
$$
\frac{d}{dt} \frac{\partial L}{\partial \dot{q_j}} -  \frac{\partial L}{\partial q_j} = 0 \text{ where } L = T-U. 
$$
For this system, this gives, after some algebra,
\begin{equation}
\begin{bmatrix} \ddot x \\ \ddot{y} \end{bmatrix} + 
\frac{h_{xx} \dot{x}^2 + 2 h_{xy} \dot{x}\dot{y} + h_{yy} \dot{y}^2 + \tilde{g}}{1+h_x^2+h_y^2} 
\begin{bmatrix}
h_x \\ h_y
\end{bmatrix}
= \begin{bmatrix}
0 \\ 0
\end{bmatrix}.
\label{xyODE}
\end{equation}
where $\tilde{g} = \tfrac{5}{7}g$.
Under the assumption that the height $h(x,y)$ varies slowly (e.g. $h(x,y)=h(\delta x,\delta y)$, $\delta\ll1$), this is approximately
\begin{equation}
\begin{split}
\ddot x + \tilde{g} h_x(x,y) &=0; \\
\ddot y + \tilde{g} h_y(x,y) & =0.
\end{split}
\label{xyODE_simp}
\end{equation}
Thus, when
\begin{equation}
\tilde{g} \ h(x,y) = U(x) + \frac{\omega^2}{2} y^2 + \epsilon F(x) y ,
\label{hxy}
\end{equation}
equation~\eqref{xyODE_simp} is identical to equation~\eqref{CCODE}. Such a surface is shown in Figure~\eqref{fig:surface}. Numerical simulations of both equations~\eqref{xyODE} and~\eqref{xyODE_simp} show qualitatively the same chaotic scattering behavior (not shown here).
Figure~\ref{fig:surface}(a) shows a contour plot of the surface used.

We denote by $x$ and $y$, respectively, the longitudinal and transverse, directions. 
To interpret the ODE, consider the case $\epsilon=0$, in which case the subspace $y = \dot{y}=0$ is invariant. A ball that begins rolling down the center of the surface with $y=0$ stays along that line. When $\epsilon>0$, this symmetry is broken, deflecting the longitudinal trajectory in the transverse direction. We will say a ``bounce'' occurs when the ball reaches a minimum in the longitudinal direction.

\subsection{Effects of dissipation}

\begin{figure}[t]
   \centering
   \includegraphics[width=\columnwidth]{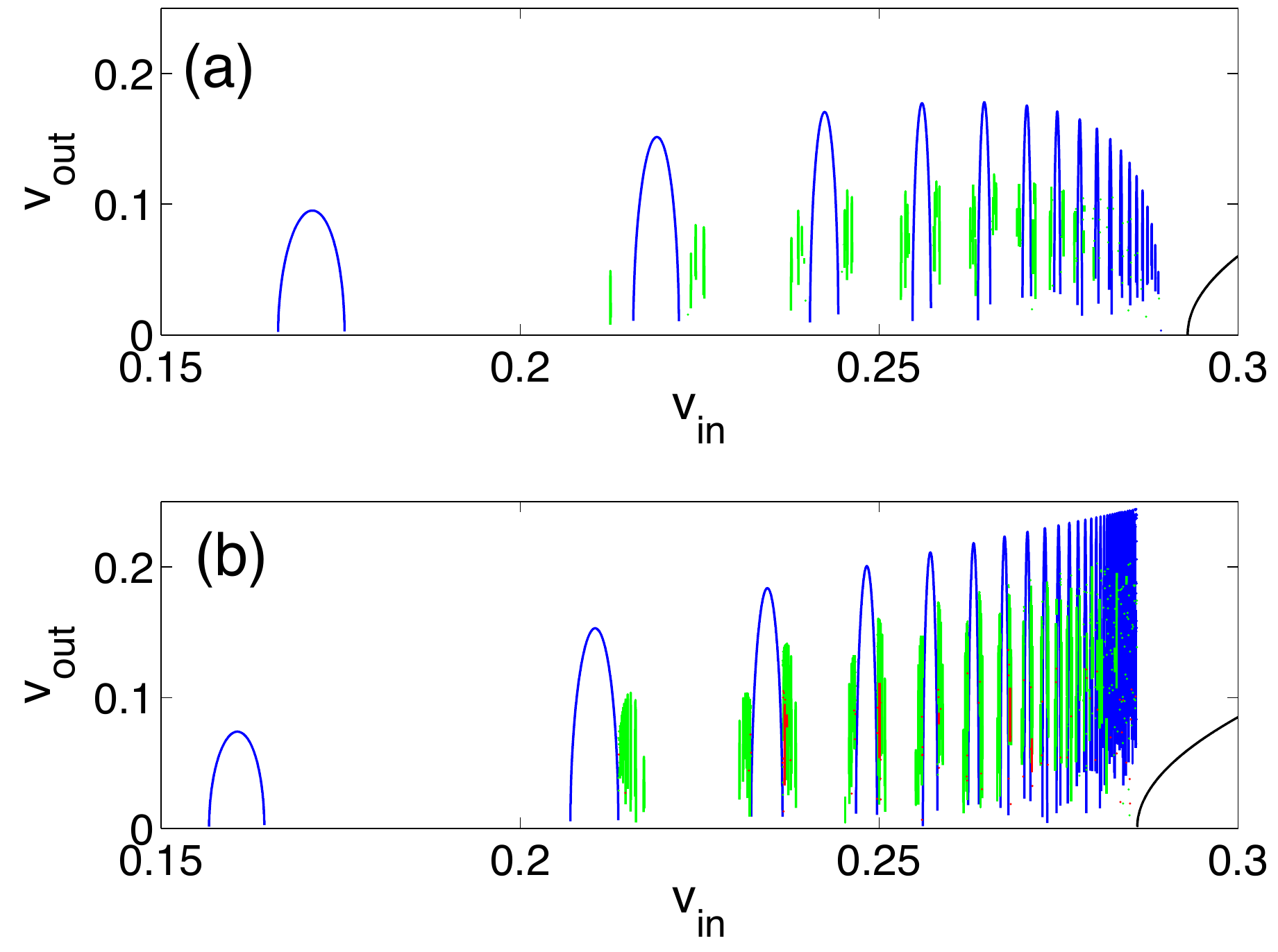} 
   \caption{Input/output plot with (a) linear damping, $\mu=0.01$ and (b) localized nonlinear damping, $\mu=0.08$.}
\label{fig:damping}
\end{figure}

To account for experimental energy loss, we include a viscous damping force $\vec{F}_{\rm damp} = -\mu (\dot x, \dot y)$, which modifies the equations of motion to:
\begin{equation*}
\begin{split}
\ddot x + \frac{5 \mu}{7 m} \dot{x} + \tilde{g} h_x(x,y) &=0; \\
\ddot y + \frac{5 \mu}{7 m} \dot{y} + \tilde{g} h_y(x,y) & =0.
\end{split}
\end{equation*}
In Goodman, Holmes, and Weinstein\cite{Goodman:2002}, a dissipative correction to system~\eqref{CCODE} is derived to account for the effect of radiative damping in nonlinear wave collisions. It adds a term like ${F(X)}^2 A^2 \dot{A}$ to the left-hand side of equation~\eqref{Addot}. This damping term is strongly localized in $X$, only applying when the kink and antikink are close together, and nonlinear in $A$ so that small radiation damps very slowly. 
Both types of dissipation alter the fractal structure displayed by the ODE system, as seen in figure~\ref{fig:damping}, but their effect is rather different. The localized damping preserves much more of the fine structure of the chaotic scattering, whereas the viscous damping destroys almost everything but the two-bounce resonance windows. Their behavior for $v$ near $\vc$ is also quite different, with viscous damping reducing the final speed of solutions in the two-bounce solutions much more than the localized damping. \emph{We therefore expect to see mostly the two-bounce resonance phenomenon in our physical experiments, rather than the full chaotic scattering picture.}

Numerical experiments were used to both verify that system~\eqref{xyODE} could reproduce the dynamics demonstrated by  solitary wave collisions. These produced the expected results, so we proceeded with laboratory experiments.

\section{The laboratory experiment}
\label{sec:experiment}

To test experimentally whether the physical system modeled by~\eqref{xyODE} displays chaotic scattering, we needed to fabricate a surface satisfying~\eqref{hxy}. The surface was  milled  out of high-density urethane foam using a three-axis mill in the Fabrication Laboratory of the NJIT Department of Architecture. Dimensions of the surface are given in Appendix~\ref{sec:appendix}. The rough milled surface was sanded and painted, as shown in Figure~\ref{fig:surface}(b).  A ramp was placed at one end, and a rubber coated steel ball (from a computer mouse), was rolled down this ramp and allowed to move along the surface until either \textbf{(1)} it became clear that it would not escape, or \textbf{(2)}~the ball returned  close enough to the starting point that we deemed it to have escaped. 

We computed an effective friction constant by the following procedure. The ball was placed at the edge of the channel near the left edge of figure~\ref{fig:surface}b and allowed to move under the influence of gravity. Coupling to the $x$-direction is weak here, so the motion remains largely confined to the $y$-direction over the time scale of observation.  From the video we find the sequence of successive maxima $\abs{y(t_k)}$ from which we can find the frequency and decay rate. Nondimensionalizing using these scales, gives $y$ motion satisfying the non dimensional equation
$$
\ddot{y} + 0.019 \ \dot{y} + y = 0.
$$
This nondimensionalization has a time unit of $0.15$ seconds. The video trials shown below last on the order of  5--10 seconds, which is 30--60 time units, over which dissipation may be significant.

In the initial experiments, we used a large marble, which lost energy rapidly. The rubber-coated steel ball, being heavier and quieter, kept rolling for much longer. Dissipation was not negligible, however, so any trial in which the ball did not exit after three ``bounces'' was considered to be trapped.

\begin{figure*}[tb] 
   \centering
   \includegraphics[width=5in]{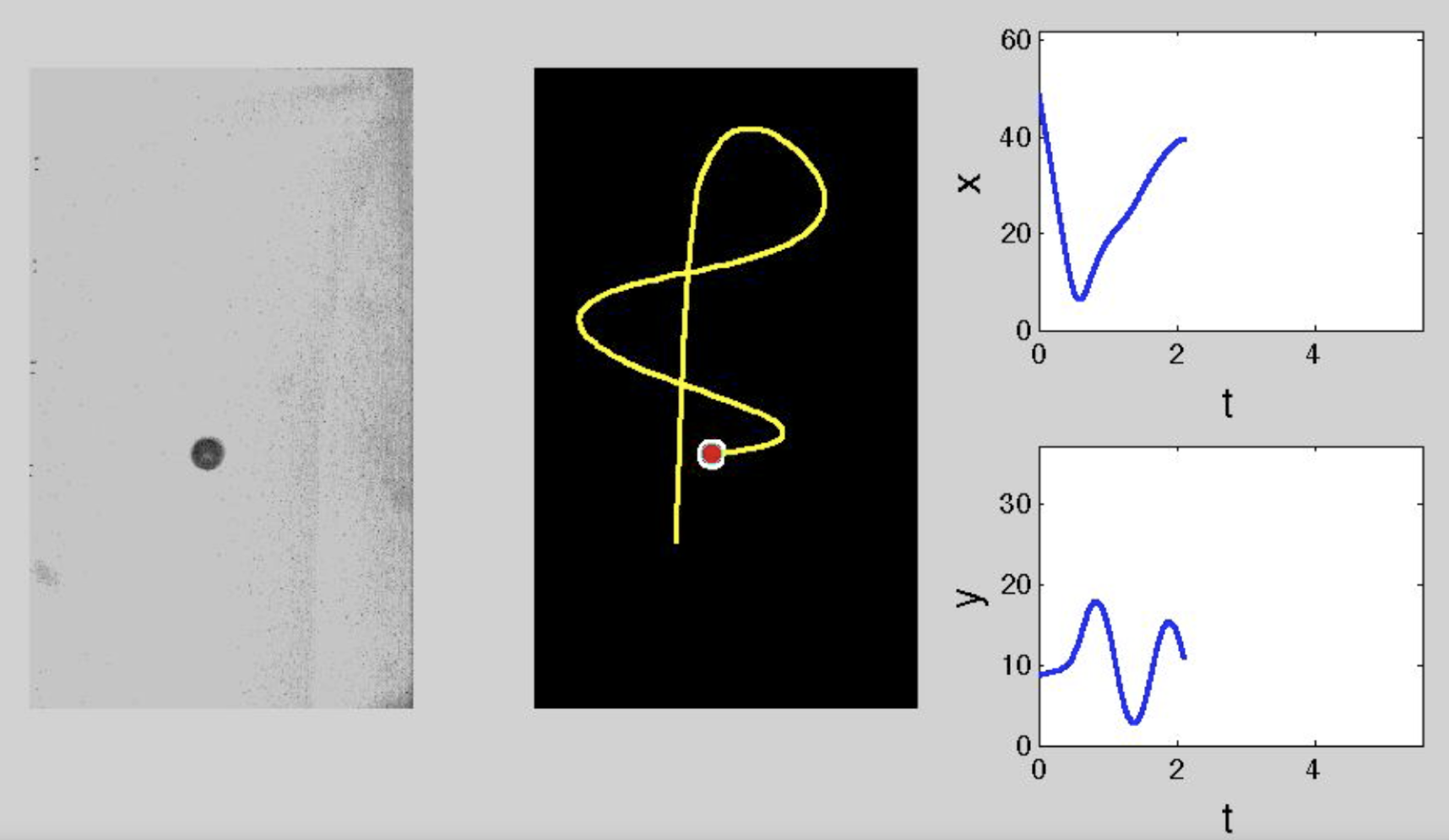} 
   \caption{(Left) One frame from a movie of the experiment. (Center) The estimated trajectory up to time $t\approx2$. (Right) The estimated coordinates as a function of time. \href{http://youtu.be/cCxCwXPYnvc}{See animation here.}}
   \label{fig:rollingballvideo}
\end{figure*}

The motion was recorded at 125 frames/second with a Photron FASTCAM 1024 PCI high-speed camera, positioned above the surface and pointing downward. Its high frame rate was not necessary, but  its short exposures prevented the ball's image from being smeared into a snake-like shape. The videos were analyzed using the MATLAB Image Processing Toolbox, giving a time series for its $x$ and $y$ coordinates. A screenshot of the video and the running program are shown in Figure~\ref{fig:rollingballvideo}.

\begin{figure}
   \centering
   \includegraphics[width=.96\columnwidth]{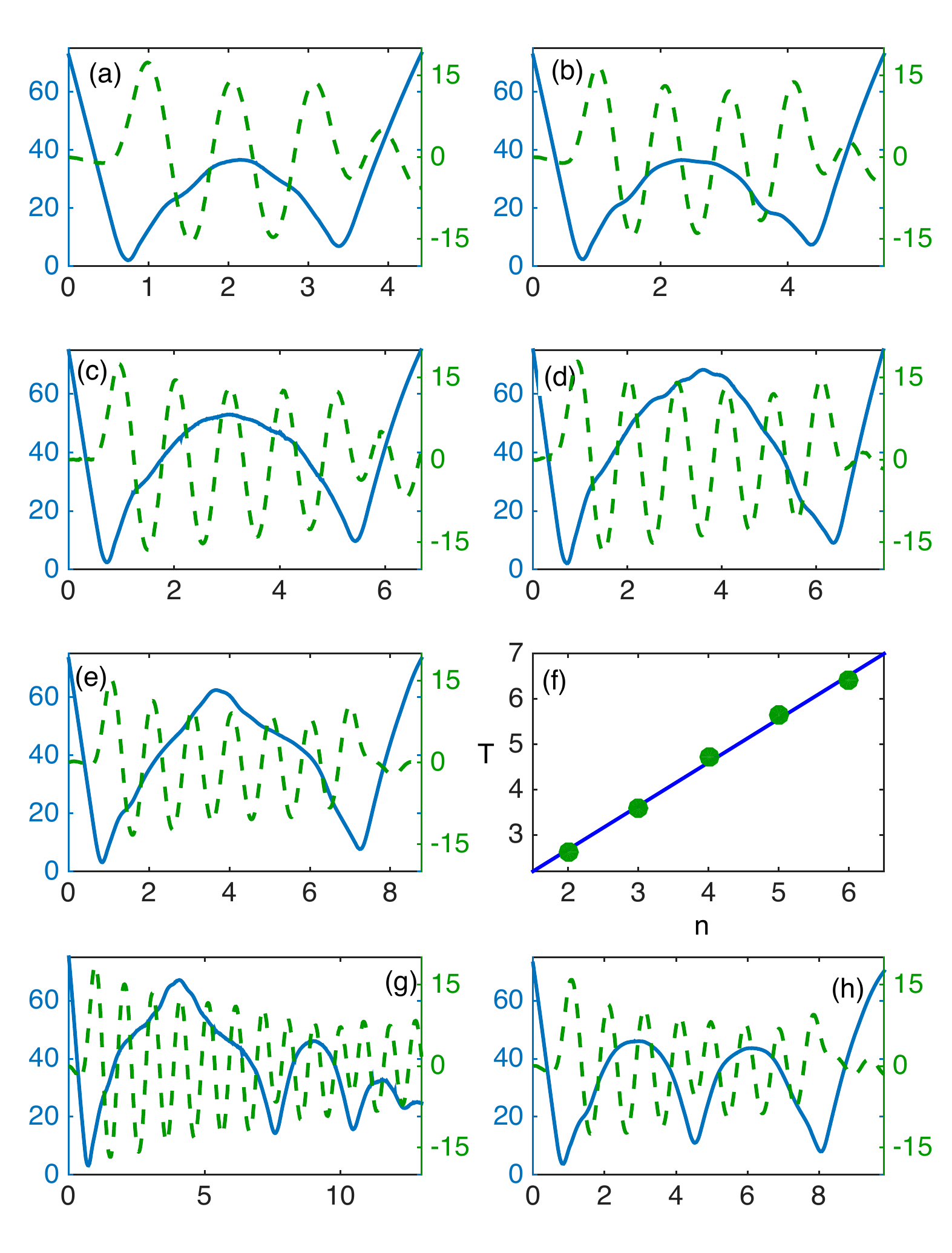}
   \caption{In all but (e) Longitudinal position $x(t)$  (solid, left $y$-axis) and transverse coordinate $y(t)$ (dashed, right $y$-axis) for selected initial velocities.(a-e)~Two bounce trajectories. Each contains one transverse oscillation more than the one that precedes it. (f)~Time between maxima of $x(t)$ versus number of transverse oscillations, with best-fit line. (g)~``Captured'' trajectory. (h)~Three bounce trajectory. }
\label{fig:July15}
\end{figure}

The initial velocity was varied by varying the height of the ball's release.
Due to the size of our foam block, the surface could only be milled to a fairly shallow depth. This meant that a ball with sufficiently large initial velocity would simply fly off the milled portion of the surface and, unfortunately, left us unable to determine the critical velocity $\vc$ experimentally. Figure~\ref{fig:July15} shows several interesting trajectories: first, five two-bounce trajectories with the number of transverse oscillations increasing from two in subfigure~(a) to six in subfigure~(e). Subfigure~(f) shows a fit of the time between the two bounces vs.\ the number of transverse oscillations, showing the same approximate linear fit as given in equation~\eqref{eq:tn}, in this case $T_n \approx  0.961\, n +   0.755$. Unfortunately, the initial velocities that led to these solutions do not increase monotonically as predicted by the theory, due, perhaps, to our inability to control with sufficient precision the transverse component of the initial location and velocity vectors. The figure also shows (g)~a trapped trajectory and (h)~a three-bounce trajectory. These represent, as far as we knox, the first (non-numerical) experimental evidence of the two-bounce resonance phenomenon, and a trace of chaotic scattering.

\section{Conclusions and Plea}
\label{sec:plea}

We have designed and built an experimental system that is approximately governed by the same reduced system that gives rise to the two-bounce resonance phenomenon in solitary wave collisions. Experiments run with this apparatus qualitatively reproduce the dynamics seen with solitary waves. 

Nonetheless, we remain hopeful that someone will find a way to demonstrate this phenomenon in a laboratory setting with actual solitary waves. The basic requirement is a solitary wave that supports an additional degree of freedom to which it can transfer energy, and which can transfer energy back. Usually, this takes the form of a mode that is localized near a stationary potential\cite{Fei:1992uu,Goodman:2004ca} or an internal mode of oscillation that moves with the wave,\cite{Campbell:1983,Peyrard:1983} although this is not always necessary.\cite{Tan:2001wi,Yang:2000wz,Dorey:2011ij}

Combing through some of the literature on this phenomenon, we find that many papers discuss the physical systems described by their equations and mention that---perhaps---this phenomenon may be found in these systems. Experimental verification for some of these systems is clearly impossible: Anninos et al.\cite{Anninos:1991vo} describe the $\phi^4$ equation as describing the interactions of large-scale domain walls in the universe. They admit, ``Because the possibility of head-on collisions is small in the real Universe, our findings are not expected to have a profound effect cosmologically.'' At the other extreme of scales, Kudryavtsev describes the $\phi^4$ system as a model for the Higgs field.\cite{Kudryavtsev:1975tc,Belova:1988} Other unpromising applications for the theory are information transport in brain microtubules,\cite{Piette:2007tm} and various other quantum field theories.\cite{Javidan:2006vo,Piette:2007tm}

Other applications are perhaps closer to being experimentally realizable. In their original paper on the subject, Campbell et al.\ suggest excitations in polymeric chains and phase transitions in uniaxial ferroelectrics.\cite{Campbell:1983} Tan and Yang study the phenomenon in vector solitons in optical fiber.\cite{Yang:2000wz,Tan:2001wi,Goodman:2005vp}  Fei et al.'s numerical experiments potentially describe the scattering of solitons off a defect in a long Josephson junction.\cite{Fei:1992uu,Goodman:2004ca} Finally, Forinash et al.\ mention the denaturation of DNA.\cite{Forinash:1994ta} This list is far from exhaustive, and we encourage readers to think if an experiment is possible in their favorite system.

\section*{Acknowledgements}
The experiment was the focus of the Spring 2010 NJIT  ``capstone'' course in applied mathematics, a class in which students perform laboratory experiments, learn the math needed to model them and apply analytical and numerical methods to that model to explain the results.  The first author was the instructor, and the others undergraduates enrolled in the  course, who continued working on the project  the following summer.  The NSF Capstone Laboratory in the NJIT Department of Mathematical Sciences was  supported by NSF DMS-0511514. Thanks to Richard Haberman, Shane Ross, and  Lawrie Virgin for useful discussions and to Richard Garber and Gene Dassing in the FabLab at the New Jersey School of Architecture, NJIT, for fabricating the surface and donating the material. Special thanks to David Campbell for early discussions about the form of this article.

\appendix
\section{Specification of the experimental apparatus}
\label{sec:appendix}

To design the experimental surface, we performed numerical experiments of system~\eqref{xyODE_simp}, varying not just the parameters in the potential~\eqref{CCODE}, but also the formulas for the potentials $U(x)$ and $F(x)$, trying to ensure we could observe interesting dynamics using the materials available to us. The surface chosen was
\begin{equation*}
h(x,y)=\eta (e^{a x}-e^{b x})+  c y^2 +  \epsilon y e^{dx}
\end{equation*}
where the units of distance in all coordinate directions is centimeters. The parameters chosen were
\begin{gather*}
a = 0.389\, \cm^{-1}, \,
b = 0.306\, \cm^{-1}, \,
c = 0.0764\, \cm^{-1}, \\
d = 0.306\, \cm^{-1}, \,
\eta = 3.27\, \cm, \,
\epsilon =  0.25.
\end{gather*}


\begin{thebibliography}{34}%
\makeatletter
\providecommand \@ifxundefined [1]{%
 \@ifx{#1\undefined}
}%
\providecommand \@ifnum [1]{%
 \ifnum #1\expandafter \@firstoftwo
 \else \expandafter \@secondoftwo
 \fi
}%
\providecommand \@ifx [1]{%
 \ifx #1\expandafter \@firstoftwo
 \else \expandafter \@secondoftwo
 \fi
}%
\providecommand \natexlab [1]{#1}%
\providecommand \enquote  [1]{``#1''}%
\providecommand \bibnamefont  [1]{#1}%
\providecommand \bibfnamefont [1]{#1}%
\providecommand \citenamefont [1]{#1}%
\providecommand \href@noop [0]{\@secondoftwo}%
\providecommand \href [0]{\begingroup \@sanitize@url \@href}%
\providecommand \@href[1]{\@@startlink{#1}\@@href}%
\providecommand \@@href[1]{\endgroup#1\@@endlink}%
\providecommand \@sanitize@url [0]{\catcode `\\12\catcode `\$12\catcode
  `\&12\catcode `\#12\catcode `\^12\catcode `\_12\catcode `\%12\relax}%
\providecommand \@@startlink[1]{}%
\providecommand \@@endlink[0]{}%
\providecommand \url  [0]{\begingroup\@sanitize@url \@url }%
\providecommand \@url [1]{\endgroup\@href {#1}{\urlprefix }}%
\providecommand \urlprefix  [0]{URL }%
\providecommand \Eprint [0]{\href }%
\providecommand \doibase [0]{http://dx.doi.org/}%
\providecommand \selectlanguage [0]{\@gobble}%
\providecommand \bibinfo  [0]{\@secondoftwo}%
\providecommand \bibfield  [0]{\@secondoftwo}%
\providecommand \translation [1]{[#1]}%
\providecommand \BibitemOpen [0]{}%
\providecommand \bibitemStop [0]{}%
\providecommand \bibitemNoStop [0]{.\EOS\space}%
\providecommand \EOS [0]{\spacefactor3000\relax}%
\providecommand \BibitemShut  [1]{\csname bibitem#1\endcsname}%
\let\auto@bib@innerbib\@empty
%</preamble>
\bibitem [{\citenamefont {Ott}\ and\ \citenamefont
  {T{\'e}l}(1993)}]{Ott:1993vs}%
  \BibitemOpen
  \bibfield  {author} {\bibinfo {author} {\bibfnamefont {E.}~\bibnamefont
  {Ott}}\ and\ \bibinfo {author} {\bibfnamefont {T.}~\bibnamefont {T{\'e}l}},\
  }\bibfield  {title} {\enquote {\bibinfo {title} {{Chaotic scattering: An
  introduction}},}\ }\href@noop {} {\bibfield  {journal} {\bibinfo  {journal}
  {Chaos}\ }\textbf {\bibinfo {volume} {3}},\ \bibinfo {pages} {417} (\bibinfo
  {year} {1993})}\BibitemShut {NoStop}%
\bibitem [{\citenamefont {Ott}(2002)}]{Ott:2002}%
  \BibitemOpen
  \bibfield  {author} {\bibinfo {author} {\bibfnamefont {E.}~\bibnamefont
  {Ott}},\ }\href {http://books.google.com/books?id=nOLx--zzHSgC} {\emph
  {\bibinfo {title} {Chaos in Dynamical Systems}}}\ (\bibinfo  {publisher}
  {Cambridge University Press},\ \bibinfo {year} {2002})\BibitemShut {NoStop}%
\bibitem [{\citenamefont {Zabusky}\ and\ \citenamefont
  {Kruskal}(1965)}]{Zabusky:1965bb}%
  \BibitemOpen
  \bibfield  {author} {\bibinfo {author} {\bibfnamefont {N.}~\bibnamefont
  {Zabusky}}\ and\ \bibinfo {author} {\bibfnamefont {M.}~\bibnamefont
  {Kruskal}},\ }\bibfield  {title} {\enquote {\bibinfo {title} {{Interaction of
  `Solitons' in a Collisionless Plasma and the Recurrence of Initial
  States}},}\ }\href@noop {} {\bibfield  {journal} {\bibinfo  {journal} {Phys.
  Rev. Lett.}\ }\textbf {\bibinfo {volume} {15}},\ \bibinfo {pages} {240--243}
  (\bibinfo {year} {1965})}\BibitemShut {NoStop}%
\bibitem [{\citenamefont {Gardner}\ \emph {et~al.}(1967)\citenamefont
  {Gardner}, \citenamefont {Greene}, \citenamefont {Kruskal},\ and\
  \citenamefont {Miura}}]{Gardner:1967vy}%
  \BibitemOpen
  \bibfield  {author} {\bibinfo {author} {\bibfnamefont {C.}~\bibnamefont
  {Gardner}}, \bibinfo {author} {\bibfnamefont {J.}~\bibnamefont {Greene}},
  \bibinfo {author} {\bibfnamefont {M.}~\bibnamefont {Kruskal}}, \ and\
  \bibinfo {author} {\bibfnamefont {R.}~\bibnamefont {Miura}},\ }\bibfield
  {title} {\enquote {\bibinfo {title} {{Method for solving the Korteweg-deVries
  equation}},}\ }\href@noop {} {\bibfield  {journal} {\bibinfo  {journal}
  {Phys. Rev. Lett.}\ }\textbf {\bibinfo {volume} {19}},\ \bibinfo {pages}
  {1095--1097} (\bibinfo {year} {1967})}\BibitemShut {NoStop}%
\bibitem [{\citenamefont {Ablowitz}\ \emph {et~al.}(1973)\citenamefont
  {Ablowitz}, \citenamefont {Kaup}, \citenamefont {Newell},\ and\ \citenamefont
  {Segur}}]{Ablowitz:1973ub}%
  \BibitemOpen
  \bibfield  {author} {\bibinfo {author} {\bibfnamefont {M.~J.}\ \bibnamefont
  {Ablowitz}}, \bibinfo {author} {\bibfnamefont {D.~J.}\ \bibnamefont {Kaup}},
  \bibinfo {author} {\bibfnamefont {A.~C.}\ \bibnamefont {Newell}}, \ and\
  \bibinfo {author} {\bibfnamefont {H.}~\bibnamefont {Segur}},\ }\bibfield
  {title} {\enquote {\bibinfo {title} {{Method for solving the sine-Gordon
  equation}},}\ }\href@noop {} {\bibfield  {journal} {\bibinfo  {journal}
  {Phys. Rev. Lett.}\ }\textbf {\bibinfo {volume} {30}},\ \bibinfo {pages}
  {1262--1264} (\bibinfo {year} {1973})}\BibitemShut {NoStop}%
\bibitem [{\citenamefont {Scott}(2004)}]{Scott:2004}%
  \BibitemOpen
  \bibfield  {author} {\bibinfo {author} {\bibfnamefont {A.}~\bibnamefont
  {Scott}},\ }\bibfield  {title} {\enquote {\bibinfo {title} {Solitons, a brief
  history},}\ }in\ \href@noop {} {\emph {\bibinfo {booktitle} {The Encyclopedia
  of Nonlinear Science}}},\ \bibinfo {editor} {edited by\ \bibinfo {editor}
  {\bibfnamefont {A.}~\bibnamefont {Scott}}}\ (\bibinfo  {publisher}
  {Routledge},\ \bibinfo {address} {New York},\ \bibinfo {year} {2004})\ pp.\
  \bibinfo {pages} {852--855}\BibitemShut {NoStop}%
\bibitem [{\citenamefont {Zabusky}\ and\ \citenamefont
  {Porter}(2010)}]{Zabusky:2010}%
  \BibitemOpen
  \bibfield  {author} {\bibinfo {author} {\bibfnamefont {N.~J.}\ \bibnamefont
  {Zabusky}}\ and\ \bibinfo {author} {\bibfnamefont {M.~A.}\ \bibnamefont
  {Porter}},\ }\bibfield  {title} {\enquote {\bibinfo {title} {Soliton},}\
  }\href@noop {} {\bibfield  {journal} {\bibinfo  {journal} {Scholarpedia}\
  }\textbf {\bibinfo {volume} {5}},\ \bibinfo {pages} {2068} (\bibinfo {year}
  {2010})}\BibitemShut {NoStop}%
\bibitem [{\citenamefont {Bondeson}, \citenamefont {Lisak},\ and\ \citenamefont
  {Anderson}(1979)}]{Bondeson:1979bh}%
  \BibitemOpen
  \bibfield  {author} {\bibinfo {author} {\bibfnamefont {A.}~\bibnamefont
  {Bondeson}}, \bibinfo {author} {\bibfnamefont {M.}~\bibnamefont {Lisak}}, \
  and\ \bibinfo {author} {\bibfnamefont {D.}~\bibnamefont {Anderson}},\
  }\bibfield  {title} {\enquote {\bibinfo {title} {{Soliton Perturbations: A
  Variational Principle for the Soliton Parameters}},}\ }\href@noop {}
  {\bibfield  {journal} {\bibinfo  {journal} {Physica Scripta}\ }\textbf
  {\bibinfo {volume} {20}},\ \bibinfo {pages} {479--485} (\bibinfo {year}
  {1979})}\BibitemShut {NoStop}%
\bibitem [{\citenamefont {Malomed}(2002)}]{Malomed:2002}%
  \BibitemOpen
  \bibfield  {author} {\bibinfo {author} {\bibfnamefont {B.~A.}\ \bibnamefont
  {Malomed}},\ }\bibfield  {title} {\enquote {\bibinfo {title} {Variational
  methods in nonlinear fiber optics and related fields},}\ }\href@noop {}
  {\bibfield  {journal} {\bibinfo  {journal} {Prog. {O}pt.}\ }\textbf {\bibinfo
  {volume} {43}},\ \bibinfo {pages} {71--193} (\bibinfo {year}
  {2002})}\BibitemShut {NoStop}%
\bibitem [{\citenamefont {Kudryavtsev}(1975)}]{Kudryavtsev:1975tc}%
  \BibitemOpen
  \bibfield  {author} {\bibinfo {author} {\bibfnamefont {A.~E.}\ \bibnamefont
  {Kudryavtsev}},\ }\bibfield  {title} {\enquote {\bibinfo {title}
  {{Solitonlike solutions for a Higgs scalar field}},}\ }\href@noop {}
  {\bibfield  {journal} {\bibinfo  {journal} {JETP Lett.}\ }\textbf {\bibinfo
  {volume} {22}},\ \bibinfo {pages} {82--83} (\bibinfo {year}
  {1975})}\BibitemShut {NoStop}%
\bibitem [{\citenamefont {Sugiyama}(1979)}]{Sugiyama:1979gl}%
  \BibitemOpen
  \bibfield  {author} {\bibinfo {author} {\bibfnamefont {T.}~\bibnamefont
  {Sugiyama}},\ }\bibfield  {title} {\enquote {\bibinfo {title} {{Kink-Antikink
  Collisions in the Two-Dimensional $\varphi^4$ Model}},}\ }\href@noop {}
  {\bibfield  {journal} {\bibinfo  {journal} {Prog. Theor. Phys.}\ }\textbf
  {\bibinfo {volume} {61}},\ \bibinfo {pages} {1550--1563} (\bibinfo {year}
  {1979})}\BibitemShut {NoStop}%
\bibitem [{\citenamefont {Ablowitz}, \citenamefont {Kruskal},\ and\
  \citenamefont {Ladik}(1979)}]{Ablowitz:1979uo}%
  \BibitemOpen
  \bibfield  {author} {\bibinfo {author} {\bibfnamefont {M.~J.}\ \bibnamefont
  {Ablowitz}}, \bibinfo {author} {\bibfnamefont {M.}~\bibnamefont {Kruskal}}, \
  and\ \bibinfo {author} {\bibfnamefont {J.}~\bibnamefont {Ladik}},\ }\bibfield
   {title} {\enquote {\bibinfo {title} {{Solitary Wave Collisions}},}\
  }\href@noop {} {\bibfield  {journal} {\bibinfo  {journal} {SIAM J. Appl.
  Math.}\ }\textbf {\bibinfo {volume} {36}},\ \bibinfo {pages} {428--437}
  (\bibinfo {year} {1979})}\BibitemShut {NoStop}%
\bibitem [{\citenamefont {Campbell}, \citenamefont {Schonfeld},\ and\
  \citenamefont {Wingate}(1983)}]{Campbell:1983}%
  \BibitemOpen
  \bibfield  {author} {\bibinfo {author} {\bibfnamefont {D.~K.}\ \bibnamefont
  {Campbell}}, \bibinfo {author} {\bibfnamefont {J.~S.}\ \bibnamefont
  {Schonfeld}}, \ and\ \bibinfo {author} {\bibfnamefont {C.~A.}\ \bibnamefont
  {Wingate}},\ }\bibfield  {title} {\enquote {\bibinfo {title} {Resonance
  structure in kink-antikink interactions in $\phi^4$ theory},}\ }\href@noop {}
  {\bibfield  {journal} {\bibinfo  {journal} {Phys. D}\ }\textbf {\bibinfo
  {volume} {9}},\ \bibinfo {pages} {1--32} (\bibinfo {year}
  {1983})}\BibitemShut {NoStop}%
\bibitem [{\citenamefont {Caputo}\ and\ \citenamefont
  {Flytzanis}(1991)}]{Caputo:1991wt}%
  \BibitemOpen
  \bibfield  {author} {\bibinfo {author} {\bibfnamefont {J.~G.}\ \bibnamefont
  {Caputo}}\ and\ \bibinfo {author} {\bibfnamefont {N.}~\bibnamefont
  {Flytzanis}},\ }\bibfield  {title} {\enquote {\bibinfo {title}
  {{Kink-antikink collisions in sine-Gordon and $\varphi^4$ models: Problems in
  the variational approach}},}\ }\href@noop {} {\bibfield  {journal} {\bibinfo
  {journal} {Phys. Rev. A}\ }\textbf {\bibinfo {volume} {44}},\ \bibinfo
  {pages} {6219} (\bibinfo {year} {1991})}\BibitemShut {NoStop}%
\bibitem [{\citenamefont {Anninos}, \citenamefont {Oliveira},\ and\
  \citenamefont {Matzner}(1991)}]{Anninos:1991vo}%
  \BibitemOpen
  \bibfield  {author} {\bibinfo {author} {\bibfnamefont {P.}~\bibnamefont
  {Anninos}}, \bibinfo {author} {\bibfnamefont {S.}~\bibnamefont {Oliveira}}, \
  and\ \bibinfo {author} {\bibfnamefont {R.}~\bibnamefont {Matzner}},\
  }\bibfield  {title} {\enquote {\bibinfo {title} {{Fractal structure in the
  scalar $\lambda (\varphi^{2}-1)^{2}$ theory}},}\ }\href@noop {} {\bibfield
  {journal} {\bibinfo  {journal} {Phys. Rev. D}\ }\textbf {\bibinfo {volume}
  {44}},\ \bibinfo {pages} {1147--1160} (\bibinfo {year} {1991})}\BibitemShut
  {NoStop}%
\bibitem [{\citenamefont {Peyrard}\ and\ \citenamefont
  {Remoissenet}(1982)}]{Peyrard:1982}%
  \BibitemOpen
  \bibfield  {author} {\bibinfo {author} {\bibfnamefont {M.}~\bibnamefont
  {Peyrard}}\ and\ \bibinfo {author} {\bibfnamefont {M.}~\bibnamefont
  {Remoissenet}},\ }\bibfield  {title} {\enquote {\bibinfo {title} {Solitonlike
  excitations in a one-dimensional atomic chain with a nonlinear deformable
  substrate potential},}\ }\href@noop {} {\bibfield  {journal} {\bibinfo
  {journal} {Phys. Rev. B}\ }\textbf {\bibinfo {volume} {26}},\ \bibinfo
  {pages} {2886--2899} (\bibinfo {year} {1982})}\BibitemShut {NoStop}%
\bibitem [{\citenamefont {Peyrard}\ and\ \citenamefont
  {Campbell}(1983)}]{Peyrard:1983}%
  \BibitemOpen
  \bibfield  {author} {\bibinfo {author} {\bibfnamefont {M.}~\bibnamefont
  {Peyrard}}\ and\ \bibinfo {author} {\bibfnamefont {D.~K.}\ \bibnamefont
  {Campbell}},\ }\bibfield  {title} {\enquote {\bibinfo {title} {Kink-antikink
  interactions in a modified sine-{G}ordon model},}\ }\href@noop {} {\bibfield
  {journal} {\bibinfo  {journal} {Physica D}\ }\textbf {\bibinfo {volume}
  {9}},\ \bibinfo {pages} {33--51} (\bibinfo {year} {1983})}\BibitemShut
  {NoStop}%
\bibitem [{\citenamefont {Remoissenet}\ and\ \citenamefont
  {Peyrard}(1984)}]{Remoissenet:1984}%
  \BibitemOpen
  \bibfield  {author} {\bibinfo {author} {\bibfnamefont {M.}~\bibnamefont
  {Remoissenet}}\ and\ \bibinfo {author} {\bibfnamefont {M.}~\bibnamefont
  {Peyrard}},\ }\bibfield  {title} {\enquote {\bibinfo {title} {Soliton
  dynamics in new models with parameterized periodic double-well and asymmetric
  substrate potentials},}\ }\href@noop {} {\bibfield  {journal} {\bibinfo
  {journal} {Phys. Rev. B}\ }\textbf {\bibinfo {volume} {29}},\ \bibinfo
  {pages} {3153--3166} (\bibinfo {year} {1984})}\BibitemShut {NoStop}%
\bibitem [{\citenamefont {Fei}, \citenamefont {Kivshar},\ and\ \citenamefont
  {Vazquez}(1992)}]{Fei:1992uu}%
  \BibitemOpen
  \bibfield  {author} {\bibinfo {author} {\bibfnamefont {Z.}~\bibnamefont
  {Fei}}, \bibinfo {author} {\bibfnamefont {Y.}~\bibnamefont {Kivshar}}, \ and\
  \bibinfo {author} {\bibfnamefont {L.}~\bibnamefont {Vazquez}},\ }\bibfield
  {title} {\enquote {\bibinfo {title} {{Resonant kink-impurity interactions in
  the sine-Gordon model}},}\ }\href@noop {} {\bibfield  {journal} {\bibinfo
  {journal} {Phys. Rev. A}\ }\textbf {\bibinfo {volume} {45}},\ \bibinfo
  {pages} {6019--6030} (\bibinfo {year} {1992})}\BibitemShut {NoStop}%
\bibitem [{\citenamefont {Zaslavsky}(2007)}]{Zaslavsky:2007}%
  \BibitemOpen
  \bibfield  {author} {\bibinfo {author} {\bibfnamefont {G.}~\bibnamefont
  {Zaslavsky}},\ }\href {http://books.google.com/books?id=W9FKkQCac8IC} {\emph
  {\bibinfo {title} {The Physics of Chaos in Hamiltonian Systems}}}\ (\bibinfo
  {publisher} {Imperial College Press},\ \bibinfo {year} {2007})\BibitemShut
  {NoStop}%
\bibitem [{\citenamefont {Gidea}, \citenamefont {de~la Llave},\ and\
  \citenamefont {Seara}(2014)}]{Gidea:2014wx}%
  \BibitemOpen
  \bibfield  {author} {\bibinfo {author} {\bibfnamefont {M.}~\bibnamefont
  {Gidea}}, \bibinfo {author} {\bibfnamefont {R.}~\bibnamefont {de~la Llave}},
  \ and\ \bibinfo {author} {\bibfnamefont {T.}~\bibnamefont {Seara}},\
  }\href@noop {} {\enquote {\bibinfo {title} {{A General Mechanism of Diffusion
  in Hamiltonian Systems: Qualitative Results}},}\ } (\bibinfo {year} {2014}),\
  \Eprint {http://arxiv.org/abs/math-ds14050866} {arXiv:math-ds14050866}
  \BibitemShut {NoStop}%
%%CITATION=MATH-DS14050866;%%
\bibitem [{\citenamefont {Goodman}\ and\ \citenamefont
  {Haberman}(2004)}]{Goodman:2004ca}%
  \BibitemOpen
  \bibfield  {author} {\bibinfo {author} {\bibfnamefont {R.~H.}\ \bibnamefont
  {Goodman}}\ and\ \bibinfo {author} {\bibfnamefont {R.}~\bibnamefont
  {Haberman}},\ }\bibfield  {title} {\enquote {\bibinfo {title} {{Interaction
  of sine-Gordon kinks with defects: the two-bounce resonance}},}\ }\href@noop
  {} {\bibfield  {journal} {\bibinfo  {journal} {Phys. D}\ }\textbf {\bibinfo
  {volume} {195}},\ \bibinfo {pages} {303--323} (\bibinfo {year}
  {2004})}\BibitemShut {NoStop}%
\bibitem [{\citenamefont {Goodman}\ and\ \citenamefont
  {Haberman}(2005{\natexlab{a}})}]{Goodman:2005vp}%
  \BibitemOpen
  \bibfield  {author} {\bibinfo {author} {\bibfnamefont {R.~H.}\ \bibnamefont
  {Goodman}}\ and\ \bibinfo {author} {\bibfnamefont {R.}~\bibnamefont
  {Haberman}},\ }\bibfield  {title} {\enquote {\bibinfo {title}
  {{Vector-soliton collision dynamics in nonlinear optical fibers}},}\
  }\href@noop {} {\bibfield  {journal} {\bibinfo  {journal} {Phys. Rev. E}\
  }\textbf {\bibinfo {volume} {71}},\ \bibinfo {pages} {56605} (\bibinfo {year}
  {2005}{\natexlab{a}})}\BibitemShut {NoStop}%
\bibitem [{\citenamefont {Goodman}\ and\ \citenamefont
  {Haberman}(2005{\natexlab{b}})}]{Goodman:2005vv}%
  \BibitemOpen
  \bibfield  {author} {\bibinfo {author} {\bibfnamefont {R.~H.}\ \bibnamefont
  {Goodman}}\ and\ \bibinfo {author} {\bibfnamefont {R.}~\bibnamefont
  {Haberman}},\ }\bibfield  {title} {\enquote {\bibinfo {title} {{Kink-antikink
  collisions in the $\varphi^4$ equation: The $n$-bounce resonance and the
  separatrix map}},}\ }\href@noop {} {\bibfield  {journal} {\bibinfo  {journal}
  {SIAM J. Appl. Dyn. Sys.}\ }\textbf {\bibinfo {volume} {4}},\ \bibinfo
  {pages} {1195--1228} (\bibinfo {year} {2005}{\natexlab{b}})}\BibitemShut
  {NoStop}%
\bibitem [{\citenamefont {Goodman}\ and\ \citenamefont
  {Haberman}(2007)}]{Goodman:2007wj}%
  \BibitemOpen
  \bibfield  {author} {\bibinfo {author} {\bibfnamefont {R.~H.}\ \bibnamefont
  {Goodman}}\ and\ \bibinfo {author} {\bibfnamefont {R.}~\bibnamefont
  {Haberman}},\ }\bibfield  {title} {\enquote {\bibinfo {title} {{Chaotic
  scattering and the $n$-bounce resonance in solitary-wave interactions}},}\
  }\href@noop {} {\bibfield  {journal} {\bibinfo  {journal} {Phys. Rev. Lett.}\
  }\textbf {\bibinfo {volume} {98}},\ \bibinfo {pages} {104103} (\bibinfo
  {year} {2007})}\BibitemShut {NoStop}%
\bibitem [{\citenamefont {Goodman}(2008)}]{Goodman:2008gk}%
  \BibitemOpen
  \bibfield  {author} {\bibinfo {author} {\bibfnamefont {R.~H.}\ \bibnamefont
  {Goodman}},\ }\bibfield  {title} {\enquote {\bibinfo {title} {{Chaotic
  scattering in solitary wave interactions: A singular iterated-map
  description}},}\ }\href@noop {} {\bibfield  {journal} {\bibinfo  {journal}
  {Chaos}\ }\textbf {\bibinfo {volume} {18}},\ \bibinfo {pages} {023113}
  (\bibinfo {year} {2008})}\BibitemShut {NoStop}%
\bibitem [{\citenamefont {Goodman}, \citenamefont {Holmes},\ and\ \citenamefont
  {Weinstein}(2002)}]{Goodman:2002}%
  \BibitemOpen
  \bibfield  {author} {\bibinfo {author} {\bibfnamefont {R.~H.}\ \bibnamefont
  {Goodman}}, \bibinfo {author} {\bibfnamefont {P.~J.}\ \bibnamefont {Holmes}},
  \ and\ \bibinfo {author} {\bibfnamefont {M.~I.}\ \bibnamefont {Weinstein}},\
  }\bibfield  {title} {\enquote {\bibinfo {title} {{Interaction of sine-Gordon
  kinks with defects: phase space transport in a two-mode model}},}\
  }\href@noop {} {\bibfield  {journal} {\bibinfo  {journal} {Phys. D}\ }\textbf
  {\bibinfo {volume} {161}},\ \bibinfo {pages} {21--44} (\bibinfo {year}
  {2002})}\BibitemShut {NoStop}%
\bibitem [{\citenamefont {Tan}\ and\ \citenamefont {Yang}(2001)}]{Tan:2001wi}%
  \BibitemOpen
  \bibfield  {author} {\bibinfo {author} {\bibfnamefont {Y.}~\bibnamefont
  {Tan}}\ and\ \bibinfo {author} {\bibfnamefont {J.}~\bibnamefont {Yang}},\
  }\bibfield  {title} {\enquote {\bibinfo {title} {{Complexity and regularity
  of vector-soliton collisions}},}\ }\href@noop {} {\bibfield  {journal}
  {\bibinfo  {journal} {Phys. Rev. E}\ }\textbf {\bibinfo {volume} {64}},\
  \bibinfo {pages} {56616} (\bibinfo {year} {2001})}\BibitemShut {NoStop}%
\bibitem [{\citenamefont {Yang}\ and\ \citenamefont {Tan}(2000)}]{Yang:2000wz}%
  \BibitemOpen
  \bibfield  {author} {\bibinfo {author} {\bibfnamefont {J.}~\bibnamefont
  {Yang}}\ and\ \bibinfo {author} {\bibfnamefont {Y.}~\bibnamefont {Tan}},\
  }\bibfield  {title} {\enquote {\bibinfo {title} {{Fractal structure in the
  collision of vector solitons}},}\ }\href@noop {} {\bibfield  {journal}
  {\bibinfo  {journal} {Phys. Rev. Lett.}\ }\textbf {\bibinfo {volume} {85}},\
  \bibinfo {pages} {3624--3627} (\bibinfo {year} {2000})}\BibitemShut {NoStop}%
\bibitem [{\citenamefont {Dorey}\ \emph {et~al.}(2011)\citenamefont {Dorey},
  \citenamefont {Mersh}, \citenamefont {Romanczukiewicz},\ and\ \citenamefont
  {Shnir}}]{Dorey:2011ij}%
  \BibitemOpen
  \bibfield  {author} {\bibinfo {author} {\bibfnamefont {P.}~\bibnamefont
  {Dorey}}, \bibinfo {author} {\bibfnamefont {K.}~\bibnamefont {Mersh}},
  \bibinfo {author} {\bibfnamefont {T.}~\bibnamefont {Romanczukiewicz}}, \ and\
  \bibinfo {author} {\bibfnamefont {Y.}~\bibnamefont {Shnir}},\ }\bibfield
  {title} {\enquote {\bibinfo {title} {{Kink-Antikink Collisions in the
  $\phi^6$ Model}},}\ }\href@noop {} {\bibfield  {journal} {\bibinfo  {journal}
  {Phys. Rev. Lett.}\ }\textbf {\bibinfo {volume} {107}} (\bibinfo {year}
  {2011})}\BibitemShut {NoStop}%
\bibitem [{\citenamefont {Belova}\ and\ \citenamefont
  {Kudryavtsev}(1988)}]{Belova:1988}%
  \BibitemOpen
  \bibfield  {author} {\bibinfo {author} {\bibfnamefont {T.~I.}\ \bibnamefont
  {Belova}}\ and\ \bibinfo {author} {\bibfnamefont {A.~E.}\ \bibnamefont
  {Kudryavtsev}},\ }\bibfield  {title} {\enquote {\bibinfo {title}
  {{Quasiperiodical Orbits In The Scalar Classical $\lambda \varphi^4$ Field
  Theory}},}\ }\href@noop {} {\bibfield  {journal} {\bibinfo  {journal} {Phys.
  D}\ }\textbf {\bibinfo {volume} {32}},\ \bibinfo {pages} {18--26} (\bibinfo
  {year} {1988})}\BibitemShut {NoStop}%
\bibitem [{\citenamefont {Piette}\ and\ \citenamefont
  {Zakrzewski}(2007)}]{Piette:2007tm}%
  \BibitemOpen
  \bibfield  {author} {\bibinfo {author} {\bibfnamefont {B.}~\bibnamefont
  {Piette}}\ and\ \bibinfo {author} {\bibfnamefont {W.}~\bibnamefont
  {Zakrzewski}},\ }\bibfield  {title} {\enquote {\bibinfo {title} {{Scattering
  of sine-Gordon kinks on potential wells}},}\ }\href@noop {} {\bibfield
  {journal} {\bibinfo  {journal} {J. Phys. A}\ }\textbf {\bibinfo {volume}
  {40}},\ \bibinfo {pages} {5995--6010} (\bibinfo {year} {2007})}\BibitemShut
  {NoStop}%
\bibitem [{\citenamefont {Javidan}(2006)}]{Javidan:2006vo}%
  \BibitemOpen
  \bibfield  {author} {\bibinfo {author} {\bibfnamefont {K.}~\bibnamefont
  {Javidan}},\ }\bibfield  {title} {\enquote {\bibinfo {title} {{Interaction of
  topological solitons with defects}},}\ }\href@noop {} {\bibfield  {journal}
  {\bibinfo  {journal} {J. Phys. A}\ }\textbf {\bibinfo {volume} {39}},\
  \bibinfo {pages} {10565--10574} (\bibinfo {year} {2006})}\BibitemShut
  {NoStop}%
\bibitem [{\citenamefont {Forinash}, \citenamefont {Peyrard},\ and\
  \citenamefont {Malomed}(1994)}]{Forinash:1994ta}%
  \BibitemOpen
  \bibfield  {author} {\bibinfo {author} {\bibfnamefont {K.}~\bibnamefont
  {Forinash}}, \bibinfo {author} {\bibfnamefont {M.}~\bibnamefont {Peyrard}}, \
  and\ \bibinfo {author} {\bibfnamefont {B.~A.}\ \bibnamefont {Malomed}},\
  }\bibfield  {title} {\enquote {\bibinfo {title} {{Interaction of discrete
  breathers with impurity modes}},}\ }\href@noop {} {\bibfield  {journal}
  {\bibinfo  {journal} {Phys. Rev. E}\ }\textbf {\bibinfo {volume} {49}},\
  \bibinfo {pages} {3400--3411} (\bibinfo {year} {1994})}\BibitemShut {NoStop}%
\end{thebibliography}
\end{document}